\documentclass
[prc,superscriptaddress,twocolumn,showpacs,preprintnu
mbers,amsmath,amssymb]{revtex4-1}
\usepackage[dvips]{graphicx}
\usepackage{dcolumn}
\usepackage{amsmath}
\usepackage{times}
\usepackage{braket}
\def\fun#1#2{\lower3.6pt\vbox{\baselineskip0pt\lineskip.9pt
 \ialign{$\mathsurround=0pt#1\hfil##\hfil$\crcr#2\crcr\sim\crcr}}}
\usepackage{graphicx}% Include figure files
\usepackage{dcolumn}% Align table columns on decimal point
\usepackage{bm}% bold math
\newcommand{\beq}{\begin{equation}}
\newcommand{\eeq}{\end{equation}}
\newcommand{\bea}{\begin{eqnarray}}
\newcommand{\eea}{\end{eqnarray}}

\newcommand{\vect}{\boldsymbol}
\usepackage{ulem}
\usepackage{color}

\begin{document}

%\preprint{APS/123-QED}

\title{Probing surface distributions of $\alpha$ clusters in $^{20}$Ne via $\alpha$-transfer reaction}


\author{Tokuro Fukui}
\email{fukui.tokuro@jaea.go.jp}
\affiliation{Nuclear Data Center, Japan Atomic Energy Agency, Tokai, Ibaraki 319-1195, Japan}

\author{Yasutaka Taniguchi}
\email{yasutaka@nims.ac.jp}
\affiliation{Department of Medical and General Sciences, Nihon Institute of Medical Science, Moroyama, Saitama 350-0435, Japan}

\author{Tadahiro Suhara}
\email{suhara@matsue-ct.jp}
\affiliation{Matsue College of Technology, Matsue, Shimane 690-8518, Japan}

\author{Yoshiko Kanada-En'yo}
\email{yenyo@ruby.scphys.kyoto-u.ac.jp}
\affiliation{Department of Physics, Kyoto University, Kyoto 606-8502, Japan}

\author{Kazuyuki Ogata}
\email{kazuyuki@rcnp.osaka-u.ac.jp}
\affiliation{Research Center for Nuclear Physics, Osaka University, Ibaraki, Osaka 567-0047, Japan}

\date{\today}% It is always \today, today,

\begin{abstract}
 \noindent
 \textbf{Background:} Direct evidence of the $\alpha$-cluster manifestation in bound states has not been obtained yet,
 although a number of experimental studies were carried out to extract the information of the clustering.
 In particular in conventional analyses of $\alpha$-transfer reactions, there exist a few significant problems on
 reaction models, which are insufficient to qualitatively discuss the cluster structure.\\
 \textbf{Purpose:} We aim to verify the development of the $\alpha$-cluster structure from observables.
 As the first application, we plan to extract the spatial information of the cluster structure of the $^{20}$Ne nucleus in its ground state
 through the cross section of the $\alpha$-transfer reaction $^{16}$O($^6$Li,~$d$)$^{20}$Ne.\\
 \textbf{Methods:} For the analysis of the transfer reaction, we work with the coupled-channel Born approximation (CCBA) approach,
 in which the breakup effect of $^6$Li is explicitly taken into account by means of the continuum-discretized coupled-channel method
 based on the three-body $\alpha + d + {}^{16}$O model.
 The two methods are adopted to calculate the overlap function between $^{20}$Ne and $\alpha + {}^{16}$O;
 one is the microscopic cluster model (MCM) with the generator coordinate method,
 and the other is the phenomenological two-body potential model (PM).\\
 \textbf{Results:} We show that the CCBA calculation with the MCM wave function gives a significant improvement of the theoretical result
 on the angular distribution of the transfer cross section, which is consistent with the experimental data.
 Employing the PM, it is discussed which region of the cluster wave function is probed on the transfer cross section.\\
 \textbf{Conclusions:} It is found that the surface region of the cluster wave function is sensitive to the cross section.
 The present work is situated as the first step in obtaining important information to systematically investigate the cluster structure.
\end{abstract}

\maketitle

%\tableofcontents

%%%%%%%%%%%%%%%%%%%%%%%%%%%%%%%%%%%%%%%%%%%%%%%%%%%%%%%%%%%%%%%%%
\section{Introduction}
\label{Intro}
%%%%%%%%%%%%%%%%%%%%%%%%%%%%%%%%%%%%%%%%%%%%%%%%%%%%%%%%%%%%%%%%%
It is common knowledge that nuclei are well described by the ``atomic like picture''
in which nucleons are considered as independent particles moving in a mean potential,
and the shell model based on this picture has achieved great success.
On the other hand, the ``molecular like picture'' can also be one with important aspects of nuclei.
It is the basic concept of the cluster model in which nucleons are regarded as strongly correlated particles forming clusters,
for example, $\alpha$ particles,
and then the clusters in nuclei weakly interact with each other.
Theoretically, the cluster structure is predicted
(for instance, in Refs.~\cite{Ikeda1980,Ohkubo1998,En'yo2001,Oertzen2006,Horiuchi2012,En'yo2012} and references therein)
to appear at the surface of not only light-stable nuclei but also $sd$-shell or unstable nuclei.
At this moment, however, there is no direct evidence from experimental studies of the nuclear cluster phenomena
except for the decay width of the resonance states.

So far, a large number of measurements of $\alpha$-transfer reactions such as ($^6$Li,~$d$), ($^7$Li,~$t$), and their inverses have been made
in order to verify the existence of $\alpha$-cluster structure.
For example, the $\alpha$-cluster structure ($\alpha+^{16}$O) of the $^{20}$Ne nucleus
was experimentally studied from the $\alpha$-transfer reaction $^{16}$O($^6$Li,~$d$)$^{20}$Ne~\cite{Becchetti1978,Anantaraman1979,Tanabe1981}
and its inverse reaction $^{20}$Ne($d$,~$^6$Li)$^{16}$O~\cite{Oelert1979} at several incident energies $E_{\rm in}$.
In their works, the $\alpha$-cluster structure of $^{20}$Ne at the $j$th excited state is discussed
by using the normalization factor $S_{j}$ in the form of the ratio $S_{j}/S_0$. 
The factor $S_{j}$ is the phenomenologically adjusted normalization one conventionally called a ``spectroscopic factor'' (SF) in
calculations with reaction models, although it is not necessarily the physical SF.
Here $j = 0$ stands for the ground state.
The reason why the relative value was used is that the absolute one is not able to be determined from the analyses
with the conventional distorted-wave Born approximation (DWBA) calculation,
some of which use a normalization factor greatly exceeding unity on the calculated cross section.
This unphysical normalization, which strongly depends on $E_{\rm in}$, is mainly due to the ambiguities of,
in the DWBA analyses~\cite{Becchetti1978,Anantaraman1979,Tanabe1981,Oelert1979},
the optical potential of $^6$Li and the $\alpha$-$^{16}$O relative wave function.
The $^6$Li optical potentials used in the DWBA analyses are not global ones, which have inconsistent parameter sets
depending on both $E_{\rm in}$ and target nuclei.
It is necessary to work with a global frame work regarding the optical potential in order to systematically investigate the $\alpha$-cluster structure.

In view of this situation, our goal is to extract the information about the spatial distribution, the surface distribution in particular, of $\alpha$ clusters from observables.
For this purpose, it is important to clarify how the $\alpha$-cluster structure is probed by the reaction.
Because the so-called SF is the inclusive quantity defined as a norm of a cluster wave function,
it is not suitable to discuss the manifestation of the cluster at the surface. 
Indeed, the SF can reach unity even if there is no spatial manifestation of clusters
because wave functions of the lowest allowed states of the SU(3) shell model are equivalent to that of the cluster model wave functions
as stated by Bayman and Bohr~\cite{Bayman1958/1959}.
This means that the SF is not appropriate for discussing the clustering phenomena.
Therefore direct comparison of calculated cross sections with measured ones is important,
and it is necessary to construct a numerically reliable theoretical framework.

In this paper, we analyze the $\alpha$-transfer reaction $^{16}$O($^6$Li,~$d$)$^{20}$Ne
in order to probe the surface distribution of the $\alpha$-cluster structure of $^{20}$Ne in its ground state
by means of the coupled-channels Born approximation (CCBA)~\cite{Penny1964,Iano1966}.
The CCBA framework is able to avoid
the aforementioned ambiguity of the $^{6}$Li optical potential by considering the three-body ($\alpha + d + {}^{16}$O) model,
in which the breakup effect of $^{6}$Li into $\alpha$ and $d$ is explicitly taken into account by employing
the method of the continuum-discretized coupled-channel (CDCC)~\cite{Kamimura1986,Austern1987,Yahiro2012}.
As the cluster model for the calculation of the $\alpha$-$^{16}$O wave function,
we adopt the microscopic cluster model (MCM) with the generator coordinate method (GCM)~\cite{Hill1953,Griffin1957,Brink1966},
which gives properties of $^{20}$Ne consistent with experimental ones.
Employing the MCM, in our framework, the aforementioned ambiguity of the $\alpha$-$^{16}$O wave function does not matter.
Through the CCBA approach with the MCM, we show a significant improvement of the 
theoretical result, which is then consistent with experimental data.
Then, in order to clarify which region of the cluster wave function is probed on the cross section,
we analyze the dependence of the cross section on the cluster wave function using 
the conventional potential model (PM), 
in which a phenomenological two-body potential is assumed as the interaction between the clusters.
%The reaction mechanisms, the breakup of $^{6}$Li and the finite-range (FR) effect in particular, are also discussed.
The breakup effect of $^{6}$Li is also discussed.

This article is outlined as follows. In Sec.~\ref{formulation} the formulation of our framework
involving both the reaction and structure models is given.
Section~\ref{modelsetting} contains the explanation of the model setting.
In Sec.~\ref{result} the result of our calculation is given.
How the $\alpha$-$^{16}$O wave function is probed on the cross section is discussed.
In Sec.~\ref{summary}, we summarize this work.

%%%%%%%%%%%%%%%%%%%%%%%%%%%%%%%%%%%%%%%%%%%%%%%%%%%%%%%%%%%%%%%%%
\section{Theoretical framework}
\label{formulation}
%%%%%%%%%%%%%%%%%%%%%%%%%%%%%%%%%%%%%%%%%%%%%%%%%%%%%%%%%%%%%%%%%
%%%%%%%%%%%%%%%%%%%%%%%%%%%%%%%%%%%%%%%%%%%%%%%%%%%%%%%%%%%%%%%%%
\subsection{Formulation of the CCBA model with CDCC}
\label{formulation1}
\begin{figure}[!t]
\begin{center}
\includegraphics[width=0.4\textwidth,clip]{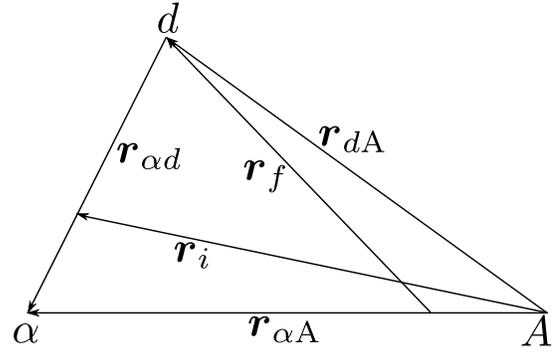}
\caption{Illustration of the three-body system.}
\label{fig1}
\end{center}
\end{figure}

%%%%%%%%%%%%%%%%%%%%%%%%%%%%%%%%%%%%%%%%%%%%%%%%%%%%%%%%%%%%%%%%%
Here, for the stripping reaction $^6{\rm Li}(\alpha+d)+A \rightarrow d+B(\alpha+A)$, we formulate the CCBA model,
in which the channel couplings regarding the continuum states of the projectile $^6{\rm Li}$ are taken into account by adopting the CDCC.
In this study we choose $^{16}$O as the target nucleus $A$, and hence the residual nucleus $B$ corresponds to $^{20}$Ne.
We assume the reaction system to be described by the three-body ($\alpha+d+A$) model shown in Fig.~\ref{fig1}.
The transition matrix ($T$ matrix) $T_{\rm CCBA}$ for the reaction is written as
\begin{align}
 T_{\rm CCBA}=
 \Braket{\Psi_f^{(-)}|V_{\rm tr}|\Psi_i^{(+)}}
 ,\label{Tmat1}
\end{align}
where $\Psi_i^{(+)}$ and $\Psi_f^{(-)}$ are the three-body wave functions in the initial and final channels, respectively,
and the transition from the former to the latter is induced by the residual interaction $V_{\rm tr}$.
Their explicit forms are given below.

The Schr\"odinger equation of the three-body wave function $\Psi_i^{(+)}$ is given by
\begin{align}
 &\left[H_i - E \right]
 \Psi_i^{(+)}(\vect{r}_{\alpha d},\vect{r}_i)
 =0
 ,\label{Scheqi}\\
 &H_i=h_i+T_{\vect{r}_i}
 +V_{\alpha \rm A}^{(\rm N)}(\vect{r}_{\alpha \rm A})+V_{dA}^{(\rm N)}(\vect{r}_{dA})
 +V_{{\rm Li}A}^{(\rm C)}(\vect{r}_i)
 ,\label{Hamili}
\end{align}
where $E$ is the total energy of the system. In the three-body Hamiltonian $H_i$, $h_i$ is the internal Hamiltonian of $^6{\rm Li}$,
and $T_{\vect{\rho}}$ is the kinetic-energy operator in relation to the coordinate $\vect{\rho}$
($\vect{\rho} = \vect{r}_{\alpha d},~\vect{r}_{\alpha A},~\vect{r}_{dA},~\vect{r}_{i},~{\rm or}~\vect{r}_{f}$).
The interaction $V_{\rm XY}$ is the nuclear or Coulomb component between particles X and Y
(${\rm X,~Y}=\alpha,~d,~A,~^6{\rm Li},~{\rm or}~B$).
Note that the superscripts (N) and (C) stand for the nuclear and Coulomb interactions, respectively.
As one can see, we disregard the Coulomb breakup, which is justified~\cite{Fukui2011}
because the effective charge of the $\alpha$-$d$ system for the electric dipole transition is almost zero.

%%%%%%%%%%%%%%%%
%%% Figure 1 %%%
%%%%%%%%%%%%%%%%

In this work we express $\Psi_i^{(+)}$ by means of the CDCC as
\begin{align}
 \Psi_i^{(+)}(\vect{r}_{\alpha d},\vect{r}_i)
 &\approx
 \sum_c
 \psi_{\alpha d}^c(\vect{r}_{\alpha d})
 \chi_i^{cc_0(+)}(\vect{r}_i)
 ,\label{CDCCi}
\end{align}
where the internal wave function $\psi_{\alpha d}^c$ of $^6{\rm Li}$ satisfies
\begin{align}
 &\left[h_i - \varepsilon_i^c \right]
 \psi_{\alpha d}^c(\vect{r}_{\alpha d})
 =0,
 \label{Scheqi2}
 \\
 &h_i=T_{\vect{r}_{\alpha d}}+V_{\alpha d}^{\rm (N+C)}\left(\vect{r}_{\alpha d}\right),
 \label{Hamili2}
\end{align}
with the energy eigenvalue $\varepsilon_i^c$.
Here the superscript (N+C) expresses the interaction containing both the nuclear and Coulomb parts.
The index $c$ represents the discretized-energy states; $c=c_0$ for the ground state, and $c \ne c_0$ for the discretized-continuum (DC) states.
In this work we disregard the intrinsic spin of $d$.
Multiplying Eq.~\eqref{Scheqi} by $\psi_{\alpha d}^{c'}$ from the left while inserting Eq.~\eqref{CDCCi} and integrating over $\vect{r}_{\alpha d}$,
we obtain the coupled-channel equation, the so-called CDCC equation~\cite{Kamimura1986,Austern1987,Yahiro2012}.
The distorted wave $\chi_i^{cc_0(+)}$ can be obtained by solving the CDCC equation under the standard boundary condition~\cite{Fukui2011}.

We adopt the residual interaction $V_{\rm tr}$ of the postform representation given by
\begin{align}
 V_{\rm tr}=V_{\alpha d}^{(\rm N+C)}(\vect{r}_{\alpha d})
           +V_{dA}^{(\rm N+C)}(\vect{r}_{dA})
           +V_{\alpha A}^{(\rm N+C)}(\vect{r}_{\alpha A})
           -U_f
 .\label{Vtr}
\end{align}
We choose the auxiliary potential $U_f$ as
\begin{align}
 U_f = V_{\alpha d}^{(\rm C)}(\vect{r}_{\alpha d})
      +V_{d A}^{(\rm N+C)}(\vect{r}_{d A})
      +V_{\alpha A}^{(\rm N+C)}(\vect{r}_{\alpha A})
 ,\label{Vaux}
\end{align}
which trivially leads to $V_{\rm tr}=V_{\alpha d}^{(\rm N)}$.

The three-body wave function $\Psi_f^{(+)}$, which is the time-reversal one of $\Psi_f^{(-)}$,
is described by $U_f$ as
\begin{align}
 &\left[H_f - E \right]
 \Psi_f^{(+)}(\vect{r}_{\alpha A},\vect{r}_f)
 =0
 ,\label{Scheqf}\\
 &H_f=T_{\vect{r}_{\alpha A}}+T_{\vect{r}_f}+U_f
 .\label{Hamilf}
\end{align}
In this paper we approximate $\Psi_f^{(+)}$ as
\begin{align}
 \Psi_f^{(+)}(\vect{r}_{\alpha A},\vect{r}_f)
 \approx
 \psi_{\alpha A}(\vect{r}_{\alpha A})
 \chi_f^{(+)}(\vect{r}_f)
 .\label{Psif1ch}
\end{align}
The distorted wave $\chi_f^{(+)}$ is generated by the effective distorting potential $\tilde{U}_f$ defined by
\begin{align}
 \tilde{U}_f = V_{dA}^{(\rm N)}(\vect{r}_f) +V_{dB}^{(\rm C)}(\vect{r}_f)
 ,\label{Uftilde}
\end{align}
which corresponds to the no-recoil limit in the final channel.
The detail of the calculation of the wave function $\psi_{\alpha A}$ describing the ground state of $B$
is given in the next section.

%%%%%%%%%%%%%%%%%%%%%%%%%%%%%%%%%%%%%%%%%%%%%%%%%%%%%%%%%%%%%%%%%
\subsection{Cluster model wave function}
\label{formulation2}
%%%%%%%%%%%%%%%%%%%%%%%%%%%%%%%%%%%%%%%%%%%%%%%%%%%%%%%%%%%%%%%%%

We define the radial part $\phi_l$ of $\psi_{\alpha A}$ with the angular momentum $l$ as
\begin{align}
 \psi_{\alpha A}\left(\vect{r}\right)
 =\phi_l\left(r\right)Y_{lm}\left(\hat{\vect{r}}\right),
 \label{phil}
\end{align}
where the coordinate $\vect{r}$ expresses the relative distance between the clusters
and $m$ is the $z$ component of the angular momentum.
In the following, how to prepare $\phi_l$ is explained.

In the MCM with the GCM, the total wave function of the two-body cluster system between $\alpha$ and $A$ is written as
\begin{align}
 \ket{\Phi_{\rm GCM}}&=
 \left|\tilde M
 \mathcal{A}\left[\phi_{l}^{\rm (GCM)}\left(r \right)Y_{l0}\left(\hat{\vect{r}}\right)\varphi_{\alpha} \varphi_A \varphi_{\rm c.m.}\right]\right>
 ,\label{GCMwf}\\
 \tilde M &\equiv \sqrt{\frac{M_{\alpha}!M_A!}{M_B!}},
 \label{tildeM}
\end{align}
where $\varphi_{\alpha}$ and $\varphi_A$ are the internal wave functions of $\alpha$ and $A$, respectively,
and the wave function $\varphi_{\rm c.m.}$ represents the motion of the total center of mass (c.m.).
The operator $\mathcal{A}$ is the antisymmetrizer which exchanges nucleons belonging to different clusters,
and $M_{\alpha}$, $M_A$, and $M_B$ stand for the mass numbers of each particle.
The wave function $\phi_{l}^{\rm (GCM)}$ contains Pauli forbidden (unphysical) states which are eliminated by 
the antisymmetrizer and does not directly equal to $\phi_{l}^{\rm (MCM)}$,
but it can be transformed to $\phi_{l}^{\rm (MCM)}$ by taking into account the antisymmetrization effect between the clusters.
Here $\phi_{l}^{\rm (MCM)}$ corresponds to $\phi_{l}$ in the MCM.
We adopt a definition of $\phi_{l}^{\rm (MCM)}$, for which $\int|\phi_{l}^{\rm (MCM)}|^2  r^2 dr=1$ .
Note that this definition of $\phi_l^{\rm (MCM)}$  is different from that in a conventional MCM, in which
the norm is usually reduced to be smaller than unity by the antisymmetrization effect.
See the Appendix for more detail.
%The difference from the conventional MCM is not the
%normalization but the definition of the function itself, though the difference is easily found in the norm.

The PM is also used to investigate in detail what the reaction probes.
The radial part $\phi_{l}^{\rm (PM)}$, which corresponds to $\phi_{l}$ in the PM,
is obtained as a solution of the Schr\"odinger equation,
\begin{align}
 \left[-\frac{\hbar^2}{2\mu_B}\left\{\frac{1}{r}\frac{d^2}{dr^2}r
 -\frac{l(l+1)}{r^2}\right\}
 +V_{\alpha A}^{(\rm N+C)}(r)\right]&
 \phi_{l}^{\rm (PM)}\left(r \right)
 \nonumber\\
 =\varepsilon_f \phi_{l}^{\rm (PM)}&\left(r \right),
 \label{Scheqf2}
\end{align}
with the energy eigenvalue $\varepsilon_f$ of $B$ in its ground state
and the reduced mass $\mu_B=M_{\alpha}M_A/\left(M_{\alpha}+M_A\right)$.
We adopt the $\alpha$-$A$ nuclear interaction $V_{\alpha A}^{(\rm N)}$ with the standard Woods-Saxon distribution
\begin{align}
 V_{\alpha A}^{(\rm N)}\left( r \right)=
 -\frac{V_0}{1+\exp\left( \frac{r-r_0}{a_0} \right)}.
 \label{WS}
\end{align}
Pauli forbidden states are eliminated by confirming the numbers of nodes of $\phi_l^{\mathrm{(PM)}}$.

It should be noted that, in general, these cluster-model wave functions, $\phi_{l}^{\rm (MCM)}$ and $\phi_{l}^{\rm (PM)}$,
are defined as
$\phi_{l}$ in Eq.~\eqref{phil} with a normalization factor:
\begin{align}
 \phi_{l}(r)=\left(S_{\rm MCM}\right)^{1/2} \phi_{l}^{\rm (MCM)}(r)
 \label{SFMCM}
\end{align}
for the MCM case, whereas for the PM case it is given by
\begin{align}
 \phi_{l}(r)=\left(S_{\rm PM}\right)^{1/2} \phi_{l}^{\rm (PM)}(r).
 \label{SFPM}
\end{align}
Here $S_{\rm MCM}$ expresses the probability of the total many-body wave function of $B$ contains 
the pure $\alpha$-$A$ cluster configuration
and is regarded as a quenching factor
because of, for example, the polarization effect of the core nucleus $A$. In principle, 
it corresponds to the physical SF, and it should be unity for an ideal $\alpha$-$A$ system without the polarization.
On the other hand, $S_{\rm PM}$ is the phenomenological normalization factor 
usually adjusted to fit the cross sections. It is not necessarily the physical SF 
but can involve an artificial renormalization factor in addition to the physical SF. 

%%%%%%%%%%%%%%%%%%%%%%%%%%%%%%%%%%%%%%%%%%%%%%%%%%%%%%%%%%%%%%%%%
\section{Model setting}
\label{modelsetting}
%%%%%%%%%%%%%%%%%%%%%%%%%%%%%%%%%%%%%%%%%%%%%%%%%%%%%%%%%%%%%%%%%
For the CDCC calculation, we use the two-range Gaussian interaction~\cite{Sakuragi1986}, labeled $V_{\alpha d}^{(\rm N)}$,
which depends on the orbital angular momentum $l_i$ between $\alpha$ and $d$.
The DC states of $^6$Li are described by employing the pseudostate method with the real-range Gaussian basis functions~\cite{Matsumoto2003}.
In this model we have the spin-degenerate resonance state of $^6$Li in the $l_i=2$ state at 2.00~MeV with a width of 0.46~MeV.
The number of Gaussian basis functions we take is 30, with a minimum (maximum) value for 1.0 (35.0)~fm of the Gaussian range parameters.
The partial waves of $\psi_{\alpha d}^c$ with respect to $l_i=0,~1,~2,~3,~{\rm and}~4$ are taken into account with up to
$\varepsilon_i^c=50,~60,~55,~60,~{\rm and}~55$~MeV, respectively.
We calculate $\psi_{\alpha d}^c$ with a maximum value of $r_{\alpha d}$ of 100.0~fm.
A uniformly charged sphere potential is used for the Coulomb interaction $V_{\alpha d}^{(\rm C)}$ with a charge radius of 3.0~ fm,
as well as for $V_{{\rm Li}A}^{(\rm C)}$ and $V_{dB}^{(\rm C)}$ with charge radii of 3.1 and 4.6~fm, respectively.
Furthermore, this Coulomb potential is also adopted for $V_{\alpha A}^{(\rm C)}$ in the PM with a charge radius of 3.1~fm.

In the calculation of the distorted wave $\chi_i^{cc_0(+)}$, we adopt
the global optical potential~\cite{Michel1983} $V_{\alpha A}^{(\rm N)}$.
On the other hand, $V_{dA}^{(\rm N)}$ is evaluated as the sum of the
proton- and neutron-optical potentials folded by the ground-state wave function
of the deuteron~\cite{Watanabe1958}. In Refs.~\cite{Watanabe2012,Watanabe2015}
it is shown that this prescription allows the three-body CDCC
based on the $d+\alpha$ two-body picture of ${}^6$Li to effectively reproduce
$^6$Li elastic cross sections calculated with the four-body CDCC
based on the $p+n+\alpha$ three-body picture of ${}^6$Li.
We take the parameter set of Dave and Gould~\cite{Dave1983}
for the nucleon global optical potential, whereas the one-range Gaussian
interaction~\cite{Ohmura1970} between $p$ and $n$ is adopted to evaluate
the deuteron wave function.
We use the deuteron global optical potential~\cite{Han2006} for $V_{dA}^{(\rm N)}$ in the final channel.

To calculate the $T$ matrix, the double integral over $r_i$ and $r_f$ is done up to 25.0~fm for both variables.
The maximum value of the total angular momentum $J$ regarding the partial waves of $\chi_i^{cc_0(+)}$ and $\chi_f^{(-)}$ is 35.
In the present calculation, the transition from the $^6$Li channels with $l_i \ne 0$ into the $d$ channel is omitted.
Note, however, that the channel couplings among all the states with $0 \le l_i\le 4$
are taken into account in solving the CDCC equation.
It is validated that the $T$-matrix elements of the transfer process from the higher partial-wave states are expected to be small~\cite{Fukui2015}
since, within the range of $V_{\alpha d}^{(\rm N)}$, the product of $V_{\alpha d}^{(\rm N)}$ and $\psi_{\alpha d}^c$ for $l_i \ne 0$
is much smaller than that for $l_i = 0$.

The GCM calculation for $\phi_{l}^{\rm (MCM)}$ of the $l=0$ ground state of $^{20}$Ne
is performed with the Volkov number~2 effective interaction of the Majorana parameter $m=0.62$~\cite{Volkov1965}
and with the width parameter $\nu=0.16~{\rm fm}^{-2}$~\cite{Matsue1975} for both $\alpha$ and $^{16}$O.
The Coulomb interaction between the clusters is explicitly taken into account by expanding it with the multirange Gaussian basis functions.
To obtain $\phi_{l}^{\rm (MCM)}$, the number of the Brink-Bloch (BB) cluster wave functions $k_{\rm max}$ is set to 10,
and we take the $\alpha$-$A$ relative distance $S_k=1,~2,~\dots,~10$ fm.
As shown in Ref.~\cite{En'yo2014}, not only the energy spectra for the ground-state band of $^{20}$Ne but also the root-mean-square radius of $^{16}$O
calculated with the present setups are consistent with the measured ones.

%%%%%%%%%%%%%%%%%%%%%%%%%%%%%%%%%%%%%%%%%%%%%%%%%%%%%%%%%%%%%%%%%
\section{Results and discussion}
\label{result}
%%%%%%%%%%%%%%%%%%%%%%%%%%%%%%%%%%%%%%%%%%%%%%%%%%%%%%%%%%%%%%%%%
%%%%%%%%%%%%%%%%%%%%%%%%%%%%%%%%%%%%%%%%%%%%%%%%%%%%%%%%%%%%%%%%%
\subsection{Result of CCBA calculation with MCM}
\label{result1}
%%%%%%%%%%%%%%%%%%%%%%%%%%%%%%%%%%%%%%%%%%%%%%%%%%%%%%%%%%%%%%%%%

%%%%%%%%%%%%%%%%
%%% Figure 2 %%%
%%%%%%%%%%%%%%%%
\begin{figure}[!b]
\begin{center}
\includegraphics[width=0.48\textwidth,clip]{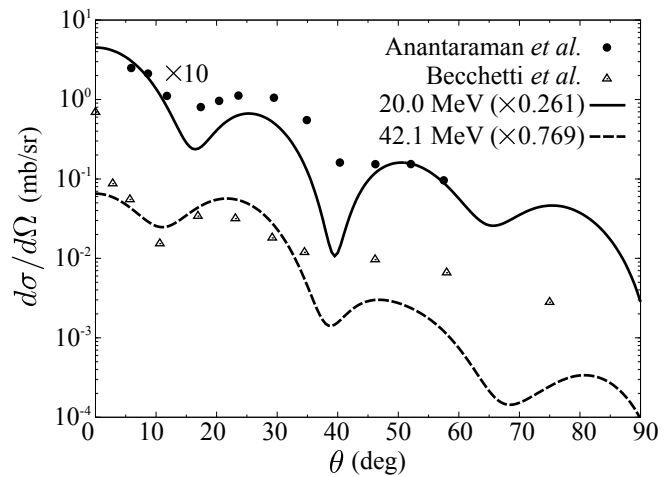}
 \caption{Calculated transfer cross section of $^{16}$O($^6$Li,~$d$)$^{20}$Ne
 at 20.0 MeV (solid line) and 42.1~MeV (dashed line)
 as a function of the deuteron emitting angle $\theta$ in the c.m. frame by
 using the MCM wave function.
 The value of $S_{\rm MCM}$, which is determined from the $\chi^2$-fit of the calculation
 to the experimental data~\cite{Becchetti1978,Anantaraman1979},
 is given in the legend.
 At 20.0~MeV the line and the dots are multiplied by 10.
 }
\label{fig2}
\end{center}
\end{figure}

We compare in Fig.~\ref{fig2} the theoretical results with the experimental data~\cite{Becchetti1978,Anantaraman1979}
for the cross section of the transfer reaction $^{16}$O($^6$Li,~$d$)$^{20}$Ne
as a function of the deuteron emitting angle $\theta$ in the c.m. frame
at $E_{\rm in}=20.0$~MeV (solid line) and 42.1~MeV (dashed line).
For the former the calculated and measured cross sections are shown after
being multiplied by 10.
The calculation is performed with the MCM wave function.
At each incident energy the line is normalized by multiplying by $S_{\rm MCM}$,
which is determined from the $\chi^2$ fit of the calculation to the experimental data
within the region $0^\circ < \theta < 80^\circ$.
The value of $S_{\rm MCM}$ is given in the legend.

One of the main consequences of this work is that our calculation improves the coincidence of the theoretical result with the experimental data
on the angular distribution of the cross section
compared to the previous DWBA analyses~\cite{Becchetti1978,Anantaraman1979} of the same reaction
as that in the present analysis.
It is remarkable that the calculation faithfully describes the diffraction pattern
around the first peak of the cross section at the angles $\theta \lesssim 10^\circ$
and the second one in the region $15^\circ \lesssim \theta \lesssim 35^\circ$.

Furthermore, we obtain the reasonable value of $S_{\rm MCM}$ of 0.261 (0.769) for $E_{\rm in}=20.0$ (42.1)~MeV
compared to the value of the normalization factor of 2.70~\cite{Anantaraman1979} (2.59~\cite{Becchetti1978})
extracted from the previous DWBA analysis,
although the energy dependence still remains.
Note that the physical SF must not be greater than unity, and therefore these factors
reported in the previous works are the phenomenological normalization factors involving artificial renormalization factors.
In Refs.~\cite{Becchetti1978,Anantaraman1979}, the $\alpha$-$^{16}$O wave function is calculated employing the PM,
i.e., the solution of Eq.~\eqref{Scheqf2} with a certain interaction $V_{\alpha A}^{\rm (N)}$ of Eq.~\eqref{WS}.
However, different parameters of $V_{\alpha A}^{(\rm N)}$ are adopted in each previous work.
Thus it is found that the appearance of the unphysical normalization factor in the DWBA analysis mainly comes from
the ambiguity of the $\alpha$-$^{16}$O wave function.
%It is expected to be accidental accidental agreement that
That two different DWBA analyses have consistent values of the phenomenological normalization factor is expected to be accidental.
The meaning of the phenomenological normalization factor in the PM  is discussed in the next section.

Another important finding is that at 42.1 MeV the DWBA calculation
employing a $^6$Li optical potential provides an unphysical
value of $S_{\rm MCM}$, even if the MCM wave function is adopted.
As mentioned above, in this work we do not need any $^6$Li optical potential.
In order to obtain a physical value of $S_{\rm MCM}$, therefore,
the description of the $^6$Li scattering based on a three-body model is found to be crucial.
%;we use the nucleon- and $\alpha$-global optical potentials.

%We reemphasize that our theoretical framework brings the bright improvement of the result
%compared to the previous DWBA analysis.
%It is due to that we avert using the $^6$Li optical potential by adopting the three-body CCBA model,
%and carry out the calculation with the reliable $\alpha$-$^{16}$O wave function
%obtained from the MCM.

%%%%%%%%%%%%%%%%%%%%%%%%%%%%%%%%%%%%%%%%%%%%%%%%%%%%%%%%%%%%%%%%%
\subsection{Discussion of the calculation with the PM}
\label{result2}
\begin{figure}[!t]
\begin{center}
\includegraphics[width=0.48\textwidth,clip]{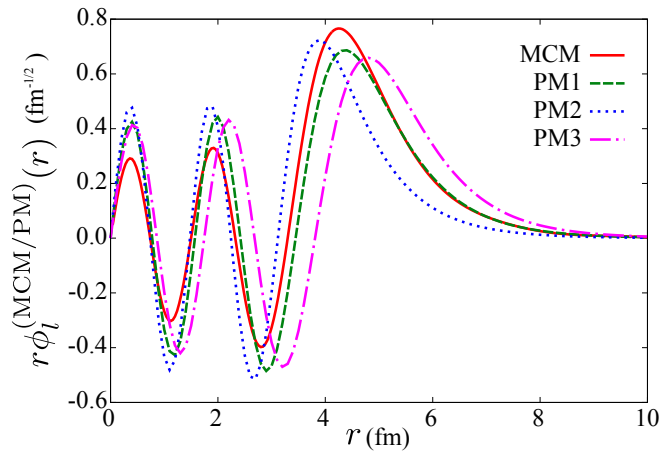}
 \caption{(Color online) The radial part of the $\alpha$-$^{16}$O relative wave function:
$\phi_l^{(\rm MCM)}$ of the MCM (solid line) and $\phi_l^{\rm (PM)}$ of 
the PM with the setups PM1 (dashed lines), PM2 (dotted line), and PM3 (dash-dotted line).
 Each wave function is normalized to have the norm one.}
\label{fig3}
\end{center}
\end{figure}

%%%%%%%%%%%%%%%%%%%%%%%%%%%%%%%%%%%%%%%%%%%%%%%%%%%%%%%%%%%%%%%%%
Here we introduce the PM in order to clarify which region of the $\alpha$-$^{16}$O wave function is probed on the transfer cross section,
and to elucidate the physical meaning of the normalization factor.
For this purpose, we prepare three types of $\phi_{l}^{(\rm PM)}$ as trial $\alpha$-$^{16}$O
wave functions
by varying the parameters $r_0$ and $a_0$ as listed in Table~\ref{table1}.
Note that the depth $V_0$ for each setup is adjusted to reproduce the $\alpha$-$^{16}$O
binding energy of 4.73~MeV.
In Fig.~\ref{fig3}, the MCM wave function $\phi_l^{\rm (MCM)}$ is shown along with $\phi_{l}^{(\rm PM)}$.
The norm of each wave function is consistently chosen to be unity.
The PM1 parameters are chosen to fit the behavior of the MCM wave function
in the tail region, say, $r~\gtrsim~5.0$~fm,
whereas the PM2 (PM3) parameters are chosen to shift the behavior to inside (outside), in particular at the surface region.

\begin{figure}[!t]
\begin{center}
\includegraphics[width=0.48\textwidth,clip]{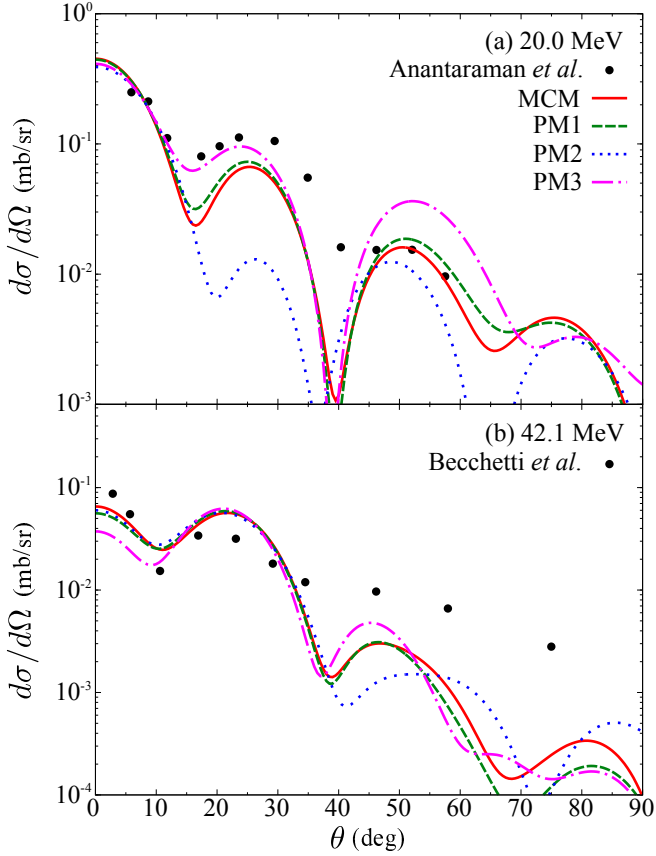}
 \caption{(Color online) Calculated cross section with the $\chi^2$ fit to be consistent with the experimental data 
 at (a)~20.0 and (b) 42.1~MeV.
 Each line corresponds to the cross section calculated with the $\alpha$-$^{16}$O wave function
 shown in Fig.~\ref{fig3}.}
\label{fig4}
\end{center}
\end{figure}

In Figs.~\ref{fig4}(a) and~\ref{fig4}(b) we show the theoretical results employing the MCM and PM wave functions
with the $\chi^2$ fit to the measured angular distribution at 20.0 and 42.1~MeV, respectively.
The factors $S_{\rm MCM}$ and $S_{\rm PM}$ extracted from the fit are listed in Table~\ref{table2}.
% In each panel, the theoretical cross sections calculated with the wave function of
% the MCM, the PM1, the PM2, and the PM3 are respectively represented by the solid, dashed, dotted, and dash-dotted lines.
At both incident energies, the MCM and the PM1 give consistent results for not only the angular distribution in Fig.~\ref{fig4}
but also $S_{\rm MCM}$ and $S_{\rm PM}$ in Table~\ref{table2}.
Therefore we can regard the results of the MCM and PM1 as nearly identical.
PM3 (PM2) at 20.0 (42.1) MeV gives an angular distribution consistent with the experimental data at the forward angles $\theta < 40^\circ$.
On the other hand, PM2 (PM3) at 20.0 (42.1) MeV underestimates the data at the second (first) peak.
Obviously, the angular distribution and $S_{\rm PM}$ depend on the PM parameters at both energies.
This fact indicates the high sensitivity of the transfer cross section to the spatial distribution of the $\alpha$-$^{16}$O relative wave function.

\begin{figure}[!t]
\begin{center}
\includegraphics[width=0.48\textwidth,clip]{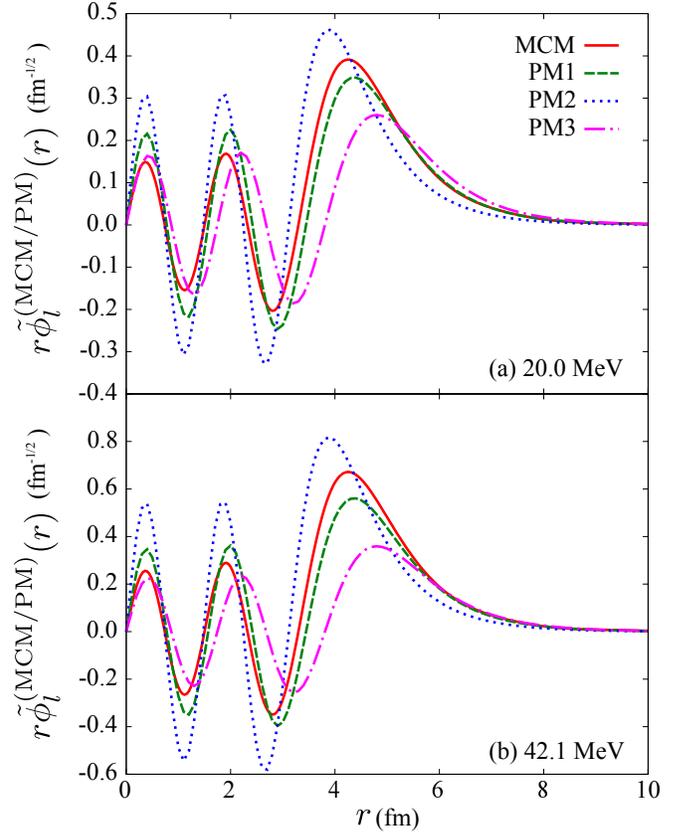}
 \caption{(Color online) $\tilde\phi_l^{\rm (MCM)}=
\left(S_{\rm MCM}\right)^{1/2}\phi_l^{\rm (MCM)}$ of
 the MCM (solid line) and 
$\tilde\phi_l^{\rm (PM)}=\left(S_{\rm PM}\right)^{1/2}\phi_l^{\rm (PM)}$ of
 PM1 (dashed line), PM2 (dotted line), and PM3 (dash-dotted line).
 The values of $S_{\rm MCM}$ and $S_{\rm PM}$ are adopted from Table~\ref{table2}
 in the case of (a) $E_{\rm in}=20.0$ and (b) 42.1~MeV.}
\label{fig5}
\end{center}
\end{figure}

Now, we introduce the wave functions
$\tilde\phi_l^{(\rm MCM)}\equiv(S_{\rm MCM})^{1/2}\phi_l^{(\rm MCM)}$ and $\tilde\phi_l^{(\rm PM)}\equiv(S_{\rm PM})^{1/2}\phi_l^{(\rm PM)}$,
where the values of $(S_{\rm MCM})^{1/2}$ and $(S_{\rm PM})^{1/2}$ are listed in Table~\ref{table2};
they behave as shown in Fig.~\ref{fig5}.
The normalization factors extracted from the transfer reaction at $E_{\rm in}=20.0$ and 42.1~MeV
are adopted in Figs.~\ref{fig5}(a) and~\ref{fig5}(b), respectively.
In each panel, the amplitudes of the PM1 and PM3 wave functions at the surface region, $r \gtrsim 6$~fm, are similar, while that of PM2 is smaller.
At $r \sim 4~\mathrm{fm}$, there is a large difference in the amplitude of each wave function.
In the following, through Fig.~\ref{fig5}, we argue we discuss what region
of $\tilde\phi_l^{(\rm PM)}$ is sensitive to the angular distribution shown in Fig.~\ref{fig4}.

When we look at Fig.~\ref{fig4}(a) for $E_{\rm in}=20.0$ MeV, 
we can see that, in terms of how well the calculation describes the behavior of the experimental cross section
at the forward angles $\theta \lesssim 40^\circ$, the results of PM1 and PM3 are consistent with the measured one.
In contrast, the diffraction pattern of the calculation with PM2 is significantly different from the experimental one.
On the other hand, in Fig.~\ref{fig5}(a), we can find that the behaviors of $\tilde\phi_l^{(\rm PM)}$ of PM1 and PM3 are similar in the region of $r \gtrsim 5$~fm,
in which PM2 has a small amplitude.
These facts indicate that the cross section at $E_{\rm in}=20.0$~MeV probes the surface region of $\tilde\phi_l^{(\rm PM)}$,
and hence its interior of $r \lesssim 5$~fm is insensitive to the cross section.

%%%%%%%%%%%%%%%
%%% Table 1 %%%
%%%%%%%%%%%%%%%
\begin{table}[!t]
\begin{center}
 \caption{The potential parameters of $V_{\alpha A}^{\rm (N)}$.
 Its depth $V_0$ is determined so as to reproduce the
 binding energy 4.73~MeV.}
\begin{tabular}{|c||c|c|}
\hline
    & $r_0$ (fm)                          & $a_0$ (fm) \\
\hline
PM1 & $1.25 \times \left(16\right)^{1/3}$ & 0.76 \\
PM2 & $1.25 \times \left(16\right)^{1/3}$ & 0.52 \\
PM3 & $1.40 \times \left(16\right)^{1/3}$ & 0.85 \\
\hline
\end{tabular}
\label{table1}
\end{center}
\end{table}

%%%%%%%%%%%%%%%%
%%% Figure 3 %%%
%%%%%%%%%%%%%%%%

At 42.1 MeV, as shown in Fig.~\ref{fig4}(b), PM2 has a
the reasonable shape, but PM3 is inconsistent with the measured distribution.
In particular, for PM3, it is remarkable that the first peak of the cross section at $\theta=0^\circ$
is significantly smaller than the second one around $\theta \sim 20^\circ$.
These can be interpreted as follows.
For PM1 and PM2, although each magnitude of $\tilde\phi_l^{(\rm PM)}$ at the surface region $r \gtrsim 4$~fm
is quite different, we find the integrated values of $\tilde\phi_l^{(\rm PM)}$ over $r$ in this region are consistent with each other.
This fact leads to the coincidence of the cross sections calculated with PM1 and PM2.
On the other hand, it can be seen that the integrated value of $\tilde\phi_l^{(\rm PM)}$ for PM3 in the region $r \gtrsim 4$~fm
is significantly smaller than that of other PM setups.
This yields the decrease of the cross section of PM3 at the forward angles.
Thus, for $E_{\rm in}=42.1$~MeV, we can conclude that the surface region $r \gtrsim 4$~fm is probed on the cross section.

%%%%%%%%%%%%%%%
%%% Table 2 %%%
%%%%%%%%%%%%%%%
\begin{table}[!t]
\begin{center}
 \caption{Normalization factor extracted from the $\chi^2$-fit of the calculated cross section.}
\begin{tabular}{|c|cccc|}
\hline
 $E_{\rm in}$ (MeV)& $S_{\rm MCM}$ & $S_{\rm PM}$ (PM1) & $S_{\rm PM}$ (PM2) &  $S_{\rm PM}$ (PM3) \\
\hline
$20.0$ & 0.261 & 0.258 & 0.407 & 0.156 \\
\hline
$42.1$ & 0.769 & 0.667 & 1.276 & 0.297 \\
\hline
\end{tabular}
\label{table2}
\end{center}
\end{table}

%%%%%%%%%%%%%%%%
%%% Figure 4 %%%
%%%%%%%%%%%%%%%%

%%%%%%%%%%%%%%%%
%%% Figure 5 %%%
%%%%%%%%%%%%%%%%

% The above discussion on the spatial sensitivity of $\tilde\phi_l^{(\rm PM)}$ to the cross section is also numerically confirmed through
% the zero-range (ZR) calculation.
% As a result of the ZR-CCBA calculation, it is found that the main strength of the integrand of the $T$ matrix exists in the region $4~\lesssim~r~\lesssim~10$~fm,
% which is consistent with the aforementioned sensitive region of $\tilde\phi_l^{(\rm PM)}$.
% It is also revealed that the absorption caused by the optical potentials in the initial channel
% is the main source of the insensitivity of the inner region of $\tilde\phi_l^{(\rm PM)}$.

Next, we clarify the physical meaning of the normalization factor.
In Fig.~\ref{fig4}(b), the results of both the PM1 and the PM2 are consistent with the experimental data.
Nevertheless, for $E_{\rm in}=42.1$~MeV, both PM parameters give inconsistent values for $S_{\rm PM}$;
one differs from the other by about a factor of 2.
Furthermore it should be especially mentioned that $S_{\rm PM}$ of PM2 exceeds unity.
When $S_{\rm PM} > 1$, it is not the physical SF, which expresses the probability of the
$\alpha$-$^{16}$O cluster configuration in $^{20}$Ne.
This unphysical value arises from the fact that, in Fig.~\ref{fig3}, the amplitude of $\phi_l^{\rm (PM)}$ of PM2
in the surface region that is sensitive to the reaction is considerably smaller than that of $\phi_l^{\rm (PM)}$
of PM1.
Consequently we need an artificial enhancement of the amplitude.
Similarly, both PM1 and PM3 describe well the angular distribution in Fig.~\ref{fig4}(a),
while $S_{\rm PM}$ of PM3 for $E_{\rm in}=20.0$~MeV is about 40\% smaller than that of PM1.
Then an artificial reduction is necessary in order to decrease the tail amplitude of $\phi_l^{\rm (PM)}$ of PM3
in Fig.~\ref{fig3}.
Hence the phenomenological normalization factor $S_{\rm PM}$, 
which is conventionally adopted in DWBA analyses employing a PM,
involves an artificial renormalization factor originating from the improper distribution of the $\alpha$-$^{16}$O wave function,
even if it has correct asymptotic behavior.
Only when we have a reliable wave function such that it gives appropriate properties for $^{20}$Ne, 
i.e., the MCM or PM1, does the normalization have a physical meaning.
It is equivalent to the SF because the normalization factor determining the tail amplitude of the MCM or PM wave function
is consistent with that of the whole amplitude.
Thus we find that the $\alpha$-transfer reaction probes not the SF but the amplitude of the cluster 
wave function at the surface region.

For a future work, a systematic analysis of the reaction at other incident energies is desired
in order to judge which $\phi_l^{\rm (PM)}$ is proper without relying on $\phi_l^{\rm (MCM)}$.
This procedure is expected to be important for the verification of the cluster structure in $sd$-shell or unstable nuclei.
Moreover, we hope this kind of systematic analysis brings us knowledge
that resolves the $E_{\rm in}$ dependence of $S_{\rm MCM}$.

%%%%%%%%%%%%%%%%%%%%%%%%%%%%%%%%%%%%%%%%%%%%%%%%%%%%%%%%%%%%%%%%%
\subsection{Break effect of $^{6}$Li}
\label{result3}
%%%%%%%%%%%%%%%%%%%%%%%%%%%%%%%%%%%%%%%%%%%%%%%%%%%%%%%%%%%%%%%%%

As mentioned above, one of the advantages of the present framework is that,
by adopting the three-body CCBA model, we can avoid using the $^6$Li optical potential,
for which there is little reliability for the $\alpha$-transfer reaction at several incident energies.
Here we show the breakup effect of $^6$Li on the cross section and discuss the applicability of the conventional DWBA
model, in which the phenomenological $^6$Li optical potential is used.

\begin{figure}[!t]
\begin{center}
\includegraphics[width=0.48\textwidth,clip]{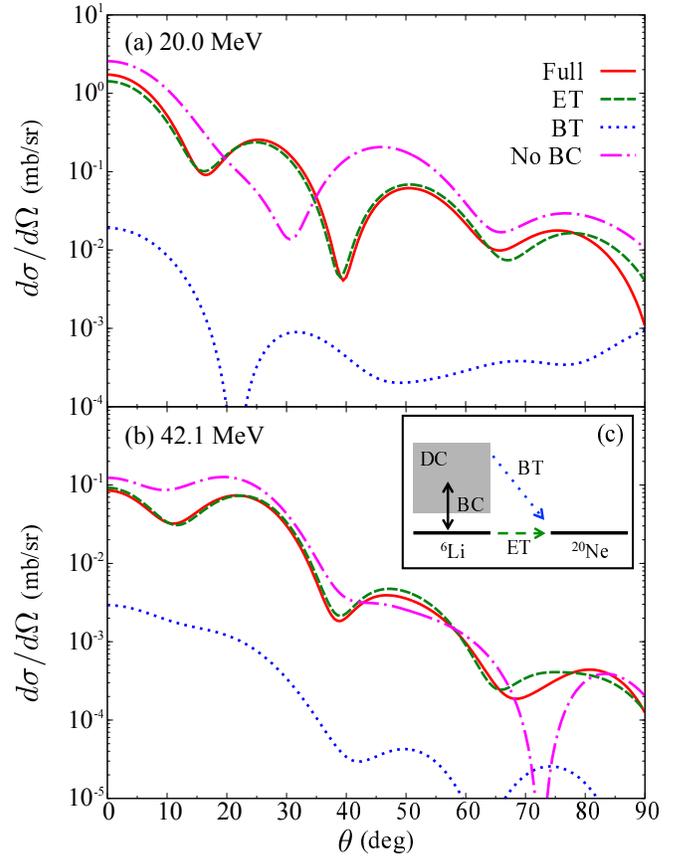}
 \caption{(Color online) Transfer processes described by the ET (dashed line) and the BT (dotted line)
 on the cross section at (a) 20.0 and (b) 42.1~MeV.
 The solid line stands for the result involving both ET and BT, while the dash-dotted line is that of the ET without the BC.
 (c) The intuitive expression of the breakup effect.
 See the text for details.}
\label{fig6}
\end{center}
\end{figure}

We decompose the $T$-matrix equation~\eqref{Tmat1} into two terms by using the CDCC wave-function equation~\eqref{CDCCi}:
 \begin{align}
  T_{\rm CCBA}&=T_{\rm ET}+T_{\rm BT}
  ,\label{Tmat2}\\
  T_{\rm ET}&=\Braket{\Psi_f^{(-)}|V_{\rm tr}|
  \psi_{\alpha d}^{c_0}(\vect{r}_{\alpha d})\chi_i^{c_0c_0(+)}(\vect{r}_i)}
 ,\label{Tmat3}\\
  T_{\rm BT}&=\Braket{\Psi_f^{(-)}|V_{\rm tr}|
  \sum_{c \ne c_0}\psi_{\alpha d}^c(\vect{r}_{\alpha d})\chi_i^{cc_0(+)}(\vect{r}_i)}
 .\label{Tmat4}
 \end{align}
$T_{\rm ET}$ describes the elastic transfer (ET),
which is the transfer process of the $\alpha$ particle from the ground state of $^6$Li to the ground state of $^{20}$Ne.
Note that the ET involves the breakup effect as the back coupling (BC), the channel couplings between the ground state
of $^6$Li and its DC states.
On the other hand, $T_{\rm BT}$ expresses the breakup transfer (BT),
in which the $\alpha$ particle transfers from the DC states to the ground state of $^{20}$Ne.
In Fig.~\ref{fig6}(c) we show the intuitive picture of these processes.

%%%%%%%%%%%%%%%%
%%% Figure 6 %%%
%%%%%%%%%%%%%%%%

Figures~\ref{fig6}(a) and~\ref{fig6}(b) show the results of the CCBA calculation with
$T_{\rm CCBA}$ (solid line), $T_{\rm ET}$ (dashed line), and $T_{\rm BT}$ (dotted line)
at $E_{\rm in}=20.0$ and 42.1~MeV, respectively.
The MCM wave function is employed for these results.
The dash-dotted line corresponds to the ET without the BC, that is, the one-step calculation.
At each incident energy, the solid and dashed lines coincide with each other in the whole region of $\theta$,
while the dash-dotted line deviates from them.
This indicates that the breakup effect of $^6$Li, the BC effect in particular, is significantly
important.
In contrast to that,
the contribution of the BT is negligibly small.
These features were reported~\cite{Fukui2011} also for the $^6$Li-induced subbarrier $\alpha$-transfer reaction~\cite{Johnson2006}.
This can intuitively be understood to be due to the hindrance of the BT by the $\alpha$-$d$ Coulomb interaction~\cite{Fukui2015}.

It should also be noted that the importance of this breakup effect does not mean the failure of the DWBA
because the BC is effectively taken into account as an ``absorption'' due to the imaginary part of the
$^6$Li optical potential.
In general, however, the optical potential is phenomenologically determined,
and it is difficult to properly evaluate the breakup effect in such a way.
Therefore the CCBA calculation should be performed to systematically
investigate the $\alpha$-cluster structure through $^6$Li-induced $\alpha$-transfer reactions.

\section{Summary}
\label{summary}
%%%%%%%%%%%%%%%%%%%%%%%%%%%%%%%%%%%%%%%%%%%%%%%%%%%%%%%%%%%%%%%%%
In order to make clear the spatial manifestation of the $\alpha$-cluster structure of $^{20}$Ne from observables,
we have analyzed the $\alpha$-transfer reaction $^{16}$O($^6$Li,~$d$)$^{20}$Ne at 20.0 and 42.1~MeV
by means of the CCBA approach based on the three-body model.
In the CCBA model the total wave function of the system in the initial channel has been described by the CDCC.
This model enables us to use the optical potentials of the subsystems $\alpha$- and $d$-$^{16}$O
instead of that of $^{6}$Li, which involves a large ambiguity.
As for the calculation of the relative wave function between the $\alpha$-$^{16}$O clusters,
the MCM with the GCM is adopted.
It is a great advantage of our procedure over the conventional approach
in which the DWBA calculation is performed with the $^{6}$Li optical potential
and a phenomenological PM for the $\alpha$-$^{16}$O wave function.

We have shown that our framework greatly improves the coincidence of the theoretical result with the experimental data
on the angular distribution of the transfer cross section.
Furthermore, we have obtained the physical value of $S_{\rm MCM}$,
for which unphysical values were reported in the previous DWBA work~\cite{Becchetti1978,Anantaraman1979}.
These improvements of the result have been brought about because
we adopted the three-body CCBA model with the reliable MCM wave function.

Next, the PM wave function has been employed in order to reveal which region of the $\alpha$-$^{16}$O wave function
is probed on the transfer cross section.
Through the comparison of the calculated cross section and the behavior of the PM wave function,
it has been concluded that the surface region of the wave function is probed on the cross section;
the region $r \gtrsim 4$~and~5~fm has been found to be sensitive to the reaction at $E_\mathrm{in} = 42.1$ and 20.0~MeV, respectively.
We have also clarified the physical meaning of the normalization factor.
As a consequence, it has been found that DWBA analyses employing a PM can have a normalization factor
involving an unphysical component originating from the improper distribution of the $\alpha$-$^{16}$O wave function,
even if it has correct asymptotic behavior.
Only when we have a reliable wave function, the normalization is equivalent to the SF.
For future work, in order to judge which $\phi_l^{\rm (PM)}$ is proper,
it is necessary to systematically analyze the reaction at several incident energies.
The present work can provide useful knowledge
for the verification of the cluster structure in $sd$-shell or unstable nuclei.

We have investigated in detail %two kinds of the reaction mechanisms characterized by the $\alpha$-$d$ system in $^6$Li.
%One is
the breakup effect of $^6$Li. %and the other is the FR effect.
It has been found to play an important role as the BC.
Since it is difficult to properly include the BC effect
in the imaginary part of the $^6$Li optical potential used in a conventional DWBA,
we need to carry out the three-body CCBA calculation for the systematic investigation of the $\alpha$-cluster structure
through $\alpha$-transfer reaction.
% As for the latter, we have shown that the ZR approximation
% works well at $E_{\rm in} = 20.0$~MeV but does not at 42.1~MeV;
% the FR calculation is necessary for the systematic analysis of
% the $\alpha$-transfer reaction.
% Clarification of the relation between the breakup effect and
% the FR effect will also be an important future work.

\appendix
%%%%%%%%%%%%%%%%%%%%%%%%%%%%%%%%%%%%%%%%%%%%%%%%%%%%%%%%%%%%%%%%%
\section{Formulation of MCM with GCM}
\label{appendix}
%%%%%%%%%%%%%%%%%%%%%%%%%%%%%%%%%%%%%%%%%%%%%%%%%%%%%%%%%%%%%%%%%
In the GCM model, the total wave function of the two-body cluster system, which consists of spinless particles $\alpha$ and $A$, can be written
as
\begin{align}
 \ket{\Phi_{\rm GCM}}=
 \sum_{k}c_k
 \ket{\Phi_{\rm BB}(S_k)}
 ,\label{GCMwf2}
\end{align}
where $\ket{\Phi_{\rm BB}(S_k)}$ is the Brink-Bloch (BB) cluster-model wave function~\cite{Brink1966} defined by
\begin{align}
 \ket{\Phi_{\rm BB}(S)}=
 \left|\tilde M
 \mathcal{A}\left\{\psi_{\alpha}\left(-\frac{M_A}{M_B}S\right)\psi_{A}\left(\frac{M_{\alpha}}{M_B}S\right)\right\}\right>
 .\label{BBwf}
\end{align}
The wave function $\psi_{\alpha}$ ($\psi_A$) of $\alpha$ $(A)$ is expressed by
the harmonic oscillator (HO) shell-model wave function with a shifted center at $\vect{S}=(0,0,S)$.
The width parameters of the HO wave functions for $\alpha$ and $A$ are assumed to be common.
The expansion coefficient $c_k$ is obtained by solving the discretized Hill-Wheeler equation
for the spin-parity eigen states projected from $\ket{\Phi_{\rm BB}(S)}$, and it is normalized  to 
satisfy $\langle \Phi_{\rm GCM} \ket{\Phi_{\rm GCM}}=1$.

Since the relative wave function between $\alpha$ and $A$ can be expressed by a localized Gaussian wave packet,
$\ket{\Phi_{\rm GCM}}$ is written as
\begin{align}
 \ket{\Phi_{\rm GCM}}&=
 \left|\tilde M
 \mathcal{A}\left[\phi_{l}^{\rm (GCM)}\left(r \right)Y_{l0}\left(\hat{\vect{r}}\right)\varphi_{\alpha} \varphi_A \varphi_{\rm c.m.}\right]\right>
 ,\label{GCMwf3}\\
 \phi_{l}^{\rm (GCM)}&\left(r\right)=
 \sum_k \sqrt{\frac{2l+1}{4\pi}} c_k
 \Gamma_l\left(r,S_k,\nu'\right)
 ,\label{relGCMwf}
\end{align}
where
the function $\Gamma_l$ is the relative wave function with its partial wave expansion~\cite{horicuhi-gcm}
defined by
\begin{align}
 \Gamma_l\left(r,S,\nu'\right) &\equiv
 4\pi \left(\frac{2\nu'}{\pi}\right)^{\frac{3}{4}}
 i_l \left(2\nu' S r\right)
 e^{-\nu' \left(r^2 +S^2 \right)}
 .\label{partialGCMwf}
\end{align}
Here $i_l$ is the modified spherical Bessel function,
and $\nu'=M_{\alpha} M_A\nu/M_B$.

By taking into account the antisymmetrization effect, we can obtain two kinds of relative wave functions
$u_l$ and $y_l$ from $\phi_{l}^{\rm (GCM)}$ as
\begin{align}
& u_l(r)=\sum_n e_n \sqrt{\mu_{2n+l}} R_{nl}(r, \nu'),\\
& y_l(r)=\sum_n e_n \mu_{2n+l} R_{nl}(r, \nu'), 
\end{align}
where $\mu_N$ $(N=2n+l)$ is the eigen value of the norm kernel, 
$R_{nl}$ is the radial part of the HO wave function,
and $e_n$ is the coefficient for the $R_{nl}$ expansion of $\phi_{l}^{\rm (GCM)}$,
\begin{align}
&\phi_{l}^{\rm (GCM)}=\sum_n e_n  R_{nl}(r, \nu').
\end{align}
Note that, in the asymptotic region where the antisymmetrization effect between clusters vanishes,
three functions, $\phi_{l}^{\rm (GCM)}$, $u_l$, and $y_l$, are identical to each other.
The wave function $u_l$ is constructed by multiplying $\phi_{l}^{\rm (GCM)}$ by the square root of the norm kernel matrix
and is a normalized wave function,
\begin{align}
\int |u_l(r)|^2 r^2 dr=1,
\end{align}
which can be regarded as a cluster wave function.
We adopt  $u_l$ as an input of the reaction calculation, $\phi_{l}^{\rm (MCM)}=u_l$.
The so-called reduced-width amplitude $y_l$ is defined as
\begin{align}
y_l(a)\equiv \frac{1}{\tilde M}
\Braket{
\frac{\delta(r-a)}{r^2}
Y_{l0}\left(\hat{\vect{r}}\right)\varphi_{\alpha} \varphi_A \varphi_{\rm c.m.}|\Phi_{\rm GCM}}
\end{align}
at a certain point $a$.
In microscopic cluster models, the so-called $S$ factor is usually defined by
the reduced-width amplitude as
\begin{align}
S=\int |y_l(r)|^2 r^2 dr.
\end{align}
The $S$ factor can be less than 1 even for the normalized GCM wave function because of the 
antisymmetrization effect. For details, the reader is referred to Ref.~\cite{horicuhi-gcm}.

%----------------------------------------------
\begin{acknowledgments}
 The authors are deeply grateful to T.~Tchuvil'sky for fruitful discussions and critical comments on our study
 and appreciate M.~Kimura and Y.~Kikuchi for their valuable comments that consolidated and clarified our argument.
 This research was supported in part by a Grant-in-Aid of the Japan Society for the Promotion
 of Science (JSPS) with Grants No.~24-2396, No.~25800124, No.~15K17662, No.~26400270, and No.~25400255.
\end{acknowledgments}
%----------------------------------------------

\nocite{*}

%+++++++++++++++++++++++++++++++++++++++++++++++++++++++++++++++++++++
%\bibliographystyle{h-physrev5TF}
%\bibliography{CIVO20Ne}

%+++++++++++++++++++++++++++++++++++++++++++++++++++++++++++++++++++++

\end{document}